\documentclass[aps,prl,twocolumn,showpacs]{revtex4}
\usepackage{graphicx}
\usepackage{dcolumn}
\usepackage{color}
\usepackage{mathrsfs}
\usepackage[breaklinks,colorlinks = true,linkcolor = blue,urlcolor=blue,citecolor=blue]{hyperref}
\usepackage{amsmath}
\usepackage{soul}
\usepackage{lipsum}
\usepackage{amssymb}
\usepackage{epstopdf}
\usepackage{subfigure}

\begin{document}
\title{Electrical and optical properties of MoS$_{2}$,MoO$_{x=2,3}$(MoSO)/RGO heterostructure}

\author{S. Erfanifam$^{1}$}
\email{s.erfanifam@gmail.com} 
\author{L. Jamilpanah$^{1}$}
\author{P. Sangpour$^{2}$}
\author{F. Haddadi$^{1}$}
\author{M. Hamdi$^{1,\footnote{Current address: Laboratory of Nanoscale Magnetic Materials and Magnonics (LMGN), Institute of Materials (IMX), School of Engineering (STI), Ecole Polytechnique Federale de Lausanne (EPFL), 1015 Lausanne, Switzerland}}$}
\author{M. Erfanifam$^{3}$}
\author{G. Chanda$^{5,6}$}
\author{T. Herrmannsd\"orfer$^{5}$}
\author{V. Sazgari$^{4}$}
\author{A. Sadeghi$^{1}$}
\author{S. Majid Mohseni$^{1}$}

\affiliation{
$^{1}$Faculty of physics, Shahid Beheshti University, Tehran, Iran\\
$^{2}$Department of Nanotechnology and Advanced Materials,Materials and Energy Research Center, Karaj, Iran\\
$^{3}$Department of physics, Zanjan univesity, Zanjan, Iran\\
$^{4}$Sabanci University Nanotechnology Research and Application Center, Tuzla, 34956 Istanbul, Turkey\\
$^{5}$Hochfeld-Magnetlabor Dresden (HLD), Helmholtz-Zentrum Dresden-Rossendorf, D-01314 Dresden, Germany\\
$^{6}$Physics department, The University of Zambia, Great east road campus, P.O. Box 32379, Lusaka, Zambia.\\
}  
\begin{abstract}
We report on transport properties of the controllable large area MoSO/Reduced graphene oxide(RGO) heterostructures electrodeposited on FTO substrates and its comparision with theoretical calculations on MoSo/Gr. I-V characteristics of the heterostructure made by P or n-type MoSO, exhibit Schottkey behavior in the interface similar to the MoS$_{2}$/Gr ones. Theoretical calculations show significant effects of lateral layer size as well as layer number in the electronic properties. In monolayer MoS$_{2}$/Gr by increasing the lateral size the energy gap disappears and the Fermi level shifts towards valence band. In the case of bilayer MoS$_{2}$ on bilayer Gr structure, the Fermi level shift is again towards valence band but, the gap is slightly higher than the monolayer structure. We found that the experimentally obtained results for n-type MoSO/RGO results are qualitatively in agreement with theoretical calculations of the MoS$_{2}$/Gr heterostructure. These  results  are  beneficial  to  understand  and  design  the  high quality and low cost MoSO/RGO based electronic, optoelectronic and energy storage devices or cocatalysts. 
\end{abstract}

\date{\today}
\maketitle
\section{Introduction}

In  the  past  decade,  low  dimensional  binary  compounds with  general  formula  MX$_{2}$ ,  has  attracted  signiﬁcant attention after discovery of graphene as a first realization of 2D materials. In these compounds M is usually a transition metal and X is a chalcogenide. The gapless characteristic of the graphene prevents its use in spintronic or optoelectronic applications while its high carrier mobility is an advantage. In contrast, the MoS$_{2}$ has a natural energy gap while the carrier mobility is low. This energy gap is not enough for room temperature applications in nanoelectronics\cite{kvashnin2018,britnel2012}. The Multilayered heterostructures in lateral or vertical shape with proper combination made of different 2D materials can overcome to these obstacles \cite{geim2013,tian2015,hsuan2017,ma2018,tan2017,huang2015,zhang2015}. Optical and electrical properties of various 2D heterostructures under gated and illumination conditions for photo response properties has been already investigated \cite{jahangir2017,huang2015,sata2015,wang2014,lili2014}. Moreover, weak Van der wasls interaction and sensitivity to the distance between these layers in performance of the fabricated electronic devices, hampers further progress \cite{zhang2017,hai2017}. Various fabrication techniques have been employed to make single or multilayer Gr and transition metal dichalcogenides (TMDs) which are mainly based on either chemical or physical growing techniques with subsequent required polymer-assissted transferring step. On the other hand polymer-assisted transfer of 2D layers can lead to contaminated layer and brings some technical difficulties. It is known that the TMDs are very brittle and difficult to grow them in a single phase and a single crystal for large areas over few hundred nanometers. Overall, the fabrication process for high quality and high cost of this methods as some of them listed above, can impede the 2D materials to enter the industrial incubators. We believe that our employed electrodeposition technique in this research can address this problems.

In this research, we report on electrical and optical properties of the MoSO/RGO electrodeposited on Fluorine doped tin oxide (FTO) substrate. In the previous report\cite{erfanifam2017}, it was observed that the oxygen content has systematic change for different thicknesses ranging from ≈ 20 to 540 nm. Optical and electrical bandgaps reveals a tunable behavior by controlling the relative content of compounds as well as a sharp transition from p to n-type of semi conductivity. Moreover, spin-orbit interaction of Mo 3d doublet enhances by reduction of MoO$_{3}$ content in thicker films. 

\section{Experiment}

The MoSO layer was electrodeposited on FTO substrate by method explained in Ref. \cite{erfanifam2017}. Growing of 2D materials in FTO substrate allows us to measure optical and electrical transport properties simultaneously. For systematic study, we chose  two p and n-type MoSO samples with 50 and 500nm thickness. Then multilayer RGO electrodeposition was carried out on MoSO layer. In this step the Whole MoSo layer was not immersed on the solution therefore we can control the overlapped region of MoSO/RGO. During the deposition, the oxygen- containing functional groups on graphene oxide sheets were reduced and deposited onto the surface of substrate. The solution used for growing the Gr layer is prepared as following.

First graphite oxide was synthesized from graphite powder by the modified hummer’s method. Briefly, 0.5g graphite powder, 0.5g NaNO$_{3}$ and 3g KMnO$_{4}$ were mixed with 23 ml H$_{2}$SO$_{4}$ solution (98 percent). The mixture was stirred in a 35  oil bath for 24h. Then 100 ml deionized (DI) water and 3 ml H$_{2}$O$_{2}$ (30 percent) were slowly added to the mixture, respectively. The color of the mixture gradually changed from dark brown to bright yellow. After that, the GO colloid solution was washed with HCl (10 percent), H$_{2}$O$_{2}$ solution and DI water three times via centrifugation, respectively. The as-prepared GO colloidal solution was diluted to a concentration of 2.5 mg/ml with DI water. 

\begin{figure}
\begin{center}
\includegraphics[scale=0.28]{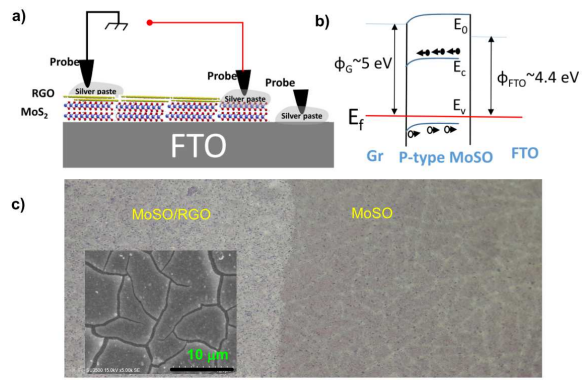}
\end{center}
\vspace{-15pt}
\caption{a: Schematic representation of the two probe measurements and b: scheme of the energy band alignment of the contacts
between graphene, P-type MoSO and FTO, c) Optical image of the MoSO and MoSO/RGO layer's interface as well as SEM micrograph. some roughness compared to the MoSO single layer in observed. However the cracks of MoSO are still present.}
\label{schim}
\end{figure}

Figure.\ref{schim}a and b, schematically represent the as grown layers and  Energy band diagrams of the proposed heterostructure. SEM micrograph in Fig.\ref{schim}c, exhibits morphology of the multilayer after RGO electrodeposition indicating a granular and uniform distribution.

\subsection{I-V characteristics and optical properties}

\begin{figure}
\begin{center}
\includegraphics[scale=0.35]{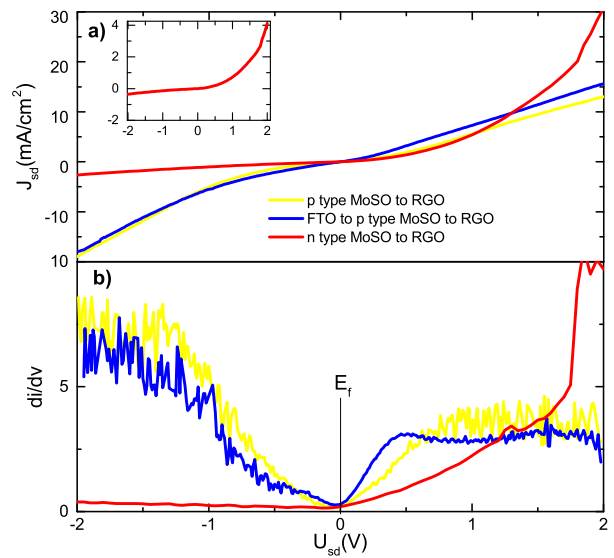}
\end{center}
\vspace{-15pt}
\caption{a: I-V characteristic curves at different probe arrangement obtained from two point measurement: one contact over heterostructure (either MoSO or RGO) and another contact on FTO substrate. b: Density of states (DOS) and Fermi level displacement calculated from first derivative of the current respect to voltage. Inset curves show an example of I-V and DOS obtained for a sample with 500nm tick MoSO. In lower panel the cutvers are shifted respect to eachother for more clarity.}
\label{iv}
\end{figure}

In order to understand the carrier transport across the schottky junctions a two probe measurement conducted for different layer thicknesses at different arrangement of the probe's position. Since in this kind of measurements the contact of the probe to surface is very important, to avoid any stress from probe to layer, the tip was immersed in the silver paste without touching the sample (see inset of Fig.\ref{schim}b). The measurement was carried out in dark condition. All measurements presented here was repeated several times at different conditions and different places in the layer. Location-independent behavior of these curves proved that in one hand our experiment is reliable and on the other hand the layer quality is high and uniform. In this series of experiments the relative change of the current is more important than the absolute values. Since, the sample is a heterostructure, the contact resistance is inevitable. The voltage sweeping rate was kept as low as 1mV/sec to be in the equilibrium condition. 

Fig.\ref{iv}a exhibits the I-V characteristic results obtained for the p and n-type MoSO in which the V$_{sd}$ is applied between RGO-MoSO, FTO-RGO and FTO-FTO. In this measurements the FTO and MoSO was grounded. The voltage was swept forward and backward (the results presented here are backward ones). The overlapped area between MoSO and RGO is about 0.575 cm$^{2}$. The asymmetric behavior of the RGO-MoSO, FTO-RGO indicates existence of a schottky interface. Detailed analysis of this curves shows that the FTO-RGO curve is more asymmetric which can be attributed to the pure perpendicular current flow and passing from double schottkey FTO/MoSO and MoSO/RGO interfaces. In contrast to that in case of MoSO-RGO the current can pass from two possible channels: 1-perpendicular with following succession: from MoSO to the FTO then from FTO to MoSO and from MoSO to RGO or 2-pass horizontally from MoSO to the interface with RGO and finally with a perpendicular current to Gr. The inset shows I-V characteristic curve of thicker MoSO layer (which we expect to be N-type semiconductor) in the MoSO-RGO mode where clearly confirms our previous work \cite{erfanifam2017}. This behavior shows a significant shift of the Fermi level toward conduction band. This data are comparable with scanning tunneling spectroscopy data. 
  
According to the Ref. \cite{cui2015} the linearity of the I-V characteristic curves at very low range of applied voltage, confirms that the, the contact resistance on Gr is very small. In addition, we believe that the observed rectification behavior can not be due to the different contact forms because the experiment was repeated for three different contacts and all resembled the same behavior.

\begin{figure}
\begin{center}
\includegraphics[scale=0.35]{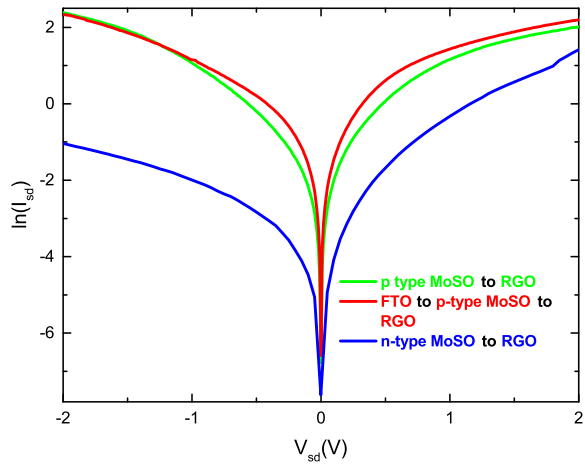}
\end{center}
\vspace{-15pt}
\caption{a: I-V characteristic curves at semi ligarithmic scale at different probe arrangement obtained from two point measurement for samples with 50 and 500nm tickness MoSO.}
\label{ideal}
\end{figure}

In positive bias electrons are transferred from RGO to MoSO and a build-in internal electric field is created from RGO to MoSO. This internal field acts as a barrier in negative bias by lowering the voltage and changing the polarity. This behavior results in band bending at MoSO/RGO interface(see Fig\ref{schim}b). 

Fig.\ref{ideal} shows the I-V characteristics in semi-log scale.  We observed the clear rectifying behavior of a p-n junction diode with low reverse current in n type MoSO/RGO and high reverse current in p type MoSO/RGO respectively. This also can imply that by increasing the MoS$_{2}$ content using thickness monitoring, the rectification parameter can be controlled. 

\begin{figure}
\begin{center}
\includegraphics[scale=0.365]{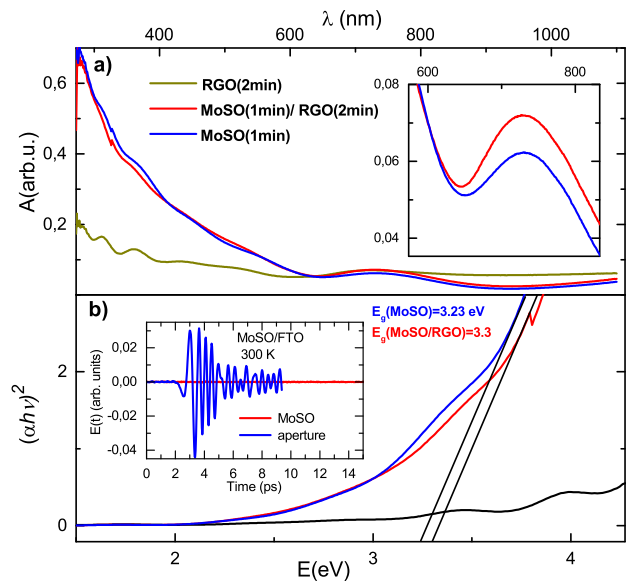}
\end{center}
\vspace{-15pt}
\caption{UV-Vis spectrum and indirect optical bandgap of MoSO/RGO structure exhibiting slightly change in the optical bandgap.  Inset shows THz measurements of the MoSO layer on FTO substrate indicating no significant change. E(t)(arb.u) is the time-dependent electric field pulse. From such a pulse we get a spectra by Fourier transformation. Dividing the Fourier transformed spectra of the sample by the spectra of open aperture, one obtains the transmission spectra shown in fig 2(a).}
\label{uv}
\end{figure}

Fig.\ref{uv} exhibits UV-Vis spectrum and optical bandgap of the MoSO/RGO heterostructure and its comparison with single MoSO and RGO layers. Optical measurements are of paramount importance in exploring phenomena related to electronic states. Transmission measurements were done with the aid of the time domain spectrometer from 1.5 eV to 4.5 eV i.e. visible to ultraviolet spectrum at 300 K. The insert of figure \ref{uv}b shows an example of one such raw data for MoSO  on FTO substrate compared to the aperture. Such raw data were also obtained for graphene on FTO and  MoSO/RGO on FTO. From such raw data visible to ultraviolet transmission spectra, shown in figure \ref{uv}a, were obtained. Transmission for the three samples generally decreases with increasing energy. However, the transmission of RGO is lower than that of MoSO and that of MoSO/RGO heterostructure at low energies pointing to the fact that conductivity of RGO is higher than that of MoSO and that of MoSO/RGO heterstructur at energies below 3eV. On the other hand, the transmission of MoSO and that of MoSO/RGO is almost similar to each other for the entire measured spectrum. 

The optical bandgap can be obtained by plotting ($\alpha$h$\nu$)$^{2}$ versus hυ (energy) (tauc plot) as shown in figure \ref{uv}b where $\alpha$ is the absorption calculated from the absorption spectrum. From the tauc plot it is clearly evident that below 3eV the curves for MoSO/RGO and MoSO coincide while at higher energy there is some difference. Further more there is a slight difference in the optical band gap MoSO (3,23 eV) compared to MoSO/RGO heterostructure (3.3 eV).  According to the gapless nature of graphene we expect no change between band gaps of MoSO and MoSO/RGO. The origin of the slight difference in optical band gaps and differences in the spectra above 3eV for the two samples (i.e. MoSO and MoSO/RGO) is not clear yet.

\section{Theoretical calculations}

The electronic properties of monolayer Graphene and has been investigated using Quantum espresso package within Perdew-Burke-Ernzerhof (PBE) approach. The projector augmented wave (PAW) pseudopotential has been used. We also take into account both spin orbit interaction and Van der Waals correction in the calculations. The kinetic energy cutoff of wavefunction and charge density are 30  and 270  respectively. First brilluine zone sampling has been carried out with a k-point mesh.
The unit cell is shown in Fig. 1 in which black balls show Carbon atoms and yellow/purple balls show S/Mo atoms. The lattice constant of Graphene and MoS$_{2}$ is 2.45  and 3.26  respectively and the vertical distance between Sulfide atoms is 3.08 It is known from previous reports \cite{hua2013} that oxygen atoms on Gr don’t lead to appearance of a significant strain.

where d$_{1}$: the distanse between Graphene layer and the layer of nearest S atoms on MoS$_{2}$. Previous works has shown the distance between Gr monolayer on MoS$_{2}$ monolayer (d$_{1}$) is 3.32A \cite{abbas2014}. This distance can define p or n type of Schottky junction \cite{liu2016} where this value 3.37A has been taken.The distance between Two MoS$_{2}$ monolayer (d$_{2}$) is 3.82A. The distance between Two Gr monolayer (d$_{3}$) is 3.3A.

\begin{figure}
\begin{center}
\includegraphics[scale=0.31]{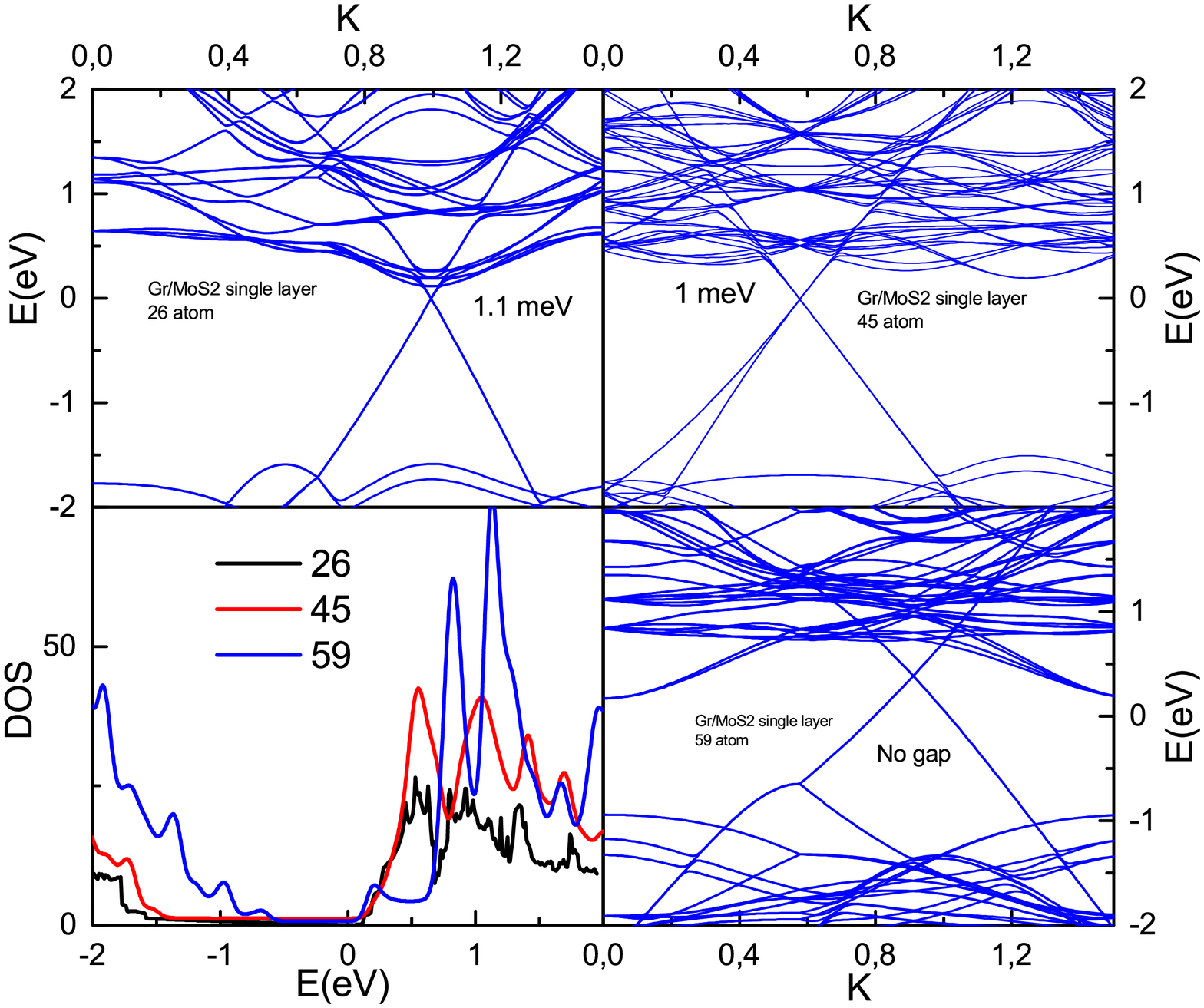}
\end{center}
\vspace{-15pt}
\caption{The calculated bandstructure of monolayer graphene on monolayer MoS$_{2}$ at different cell sizes including 26, 45 and 59 atoms. The related DOS with some systematic change  below and abow Fermi level is shown.}
\label{grmossingltheo}
\end{figure}

Fig.\ref{grmossingltheo} exhibits bandstructure calculations and DOS of the monolayer MoS$_{2}$ on monolayer Gr for different cell sizes (26, 45 and 59 atoms). Interestingly, we observed that by increasing the cell size the bandgap is going to be disappear and density of states in conduction band is increasing. However, the Fermi level is shifting to the valence band in agreement with our experimental results. In addition we observed that the Schottky barrier, which is required energy to excite electrons from graphene to MoS$_{2}$, is rising from 116meV for 26 atom cell to 730 meV in 59 atom cell.
Our measured DOS and Schottky barrier are consistent with the reported theoretical calculations \cite{hieu2017,liu2015}.

Calculations based on the cell with bilayer graphene and MoS$_{2}$ exhibits double (and may be triple) Dirac cones with massless Dirac Fermions due to the signiﬁcant charge transfer between the graphene plane and MoS$_{2}$ that enhances some energetic stability. The Schottky barrier value is lowering in double layer heterostructure from 520 meV to 87 meV by increasing the cell size. In addition, the Ref. \cite{hua2013} show 39meV bandgap (for an smaller cell size) compared to our calculations (with bigger cell size and no oxygen content) which is in the range of E$_{g}$=0 to E$_{g}$=5.9 meV. 

\begin{figure}
\begin{center}
\includegraphics[scale=0.32]{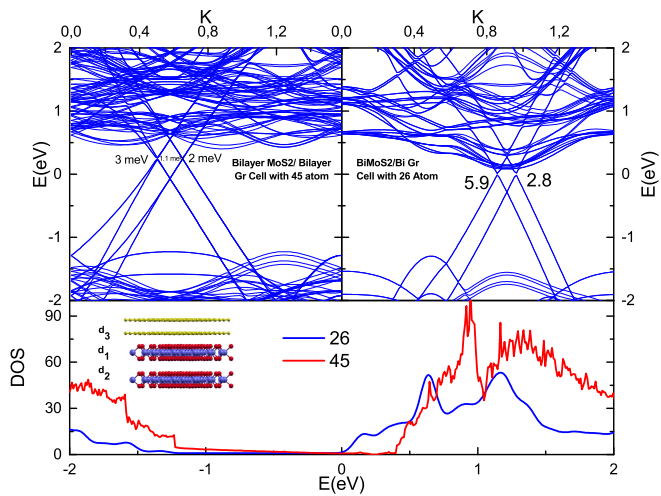}
\end{center}
\vspace{-15pt}
\caption{The calculated bandstructure of bilayer graphene on bilayer MoS$_{2}$ at different cell sizes including 26 and 45 atoms. The related DOS with some systematic change below and above Fermi level is shown.}
\label{fft}
\end{figure}

\begin{table}
\caption{Calculated $MoS_{2}/Gr$ multilayer properties.}
\vspace{.25cm}
\centering
\begin{tabular}{c c c c c c c c}
\hline\hline
Cell & Strain & a({Gr})(A$^{0}$) & a({MoS$_{2}$}) & d$_{1}$ & d$_{2}$ & d$_{3}$ & E$_{g}(eV)$\\ [0.8ex]
\hline 
$26$ & $-2.93$ \% on Gr & $2.388$ & $3.16$ & $3.33$ &  &  & 1.1\\
\hline
45 & -1.93 \% on Gr & 2.413 & 3.16 & 3.3 &  &  & 1\\ 
\hline
59 & 3.2 \% on MoS$_{2}$ & 2.45 & 3.26 & 3.38 &  &  & 0\\
\hline
52 & -2.93 \% on Gr & 3.388 & 3.16 & 3.33 & 3.74 & 3.42 & 5.9\\
   &                &       &      &      &      &      & 2.8\\
\hline
90 & -1.93 \% on Gr & 2.413 & 3.16 & 3.33 & 3.77 & 3.39 & 3\\
   &                &       &      &      &      &      & 1.1\\
	 &                &       &      &      &      &      & 2\\
 [2ex]
\hline
\end{tabular}
\label{relative}
\end{table}


\section{Conclusion}

In conclusion the electrical transport properties the MoSO/RGO heterojunction shows a Schottky behavior similar to MoS$_{2}$/Gr indicating insensitivity to the oxygen content. The type of the schottky barrier is modulated by this oxygen content instead of applying gate voltage or changing the interfacial distance which make it a reproducible and reliable technique for device fabrication.
In addition, surprisingly, we observed that the RGO layer does not changes (or at least very insensible) the bandgap but increases the adhesion of the heterostructure to substrate which increases the stability against mechanical damages or chemical corrosion suitable for device fabrication. Tunable Schottky barrier arising from tunable bandgap of the MoSO, in this heterostructure is another prominant result. 
The theoretical calculation indicated that by increasing the atomic cell size or layer number the gap can be reduced in agreement with experiment. Theoretical calculations presented here for a heterostructure with no Oxygen content can be developed for MoSo/Gr which should be addressed in future. More investigations on patterned/non patterned form of this heterostructure for further electronic properties (such as carrier mobility)or solution based experiments for electrochemical properties applicable as electrodes,  at different conditions are suggested for future researches.

\begin{acknowledgments}
S. Erfanifam acknowledges the support from the Iranian Elite's Foundation.
\end{acknowledgments}



\begin{thebibliography}{99}

\bibliographystyle{usrt}
\bibitem{britnel2012}L. Britnell, R. V. Gorbachev, R. Jalil, B. D. Belle, F. Schedin, A. Mishchenko, T. Georgiou, M. I. Katsnelson, L. Eaves, S. V. Morozov, N. M. R. Peres, J. Leist, A. K. Geim, K. S. Novoselov, L. A. Ponomarenko, "Field-Effect Tunneling Transistor Based on Vertical Graphene Heterostructures", Science, 2012, 335, 947-950.
\bibitem{kvashnin2018} A. G. Kvashnin, P. B. Sorokin, Leonid A. Chernozatonskii, Computational Materials Science, 2018, 142, 32–37.
\bibitem{geim2013}A. K. Geim and I. V. Grigorieva, "Van der Waals heterostructures", Nature, 2013, 499, 420.
\bibitem{tian2015}Xiaoqing Tian, Lin Liu, Yu Du, Juan Gu, Jian-bin Xu, and Boris I. Yakobson,  Phys.Chem.Chem.Phys.,"Variable electronic properties of lateral phosphorene–graphene heterostructures"2015, 17, 31685.
\bibitem{hsuan2017}Hsuan-An  Chen,  Wei-Chan  Chen,  Hsu  Sun,  Chien-Chung  Lin and Shih-Yen Lin, "Scalable MoS$_{2}$ /Graphene Heterostructures Grown Epitaxially on Sapphire Substrates for Phototransistor Applications", Semicond. Sci. Technol. in press 2017.
\bibitem{ma2018}Jun Ma, He Bai, Wei Zhao, Yujie Yuan, Kailiang Zhang, "High efficiency graphene/MoS$_{2}$/Si Schottky barrier solar cells using layer-controlled MoS$_{2}$ ﬁlms", Solar Energy, 2018, 160, 76–84.
\bibitem{tan2017} Haijie Tan, Wenshuo Xu, Yuewen Sheng, Chit Siong Lau, Ye Fan, Qu Chen, Martin Tweedie, Xiaochen Wang, Yingqiu Zhou, and Jamie H. Warner, Adv. Mater. 2017, 1702917.
\bibitem{zhang2015}Wenjing Zhang, Qixing Wang, Yu Chen, Zhuo Wang, and Andrew T S Wee, "Van der Waals stacked 2D layered materials for optoelectronics", 2D Mater., 2016, 3, 022001.
\bibitem{huang2015} Zongyu Huang, Weijia Han, Hongli Tang, Long Ren, D Sathish Chander, Xiang Qi, and Han Zhang, "Photoelectrochemical-type sunlight photodetector based on MoS$_{2}$/graphene heterostructure", 2D Mater., 2015, 2, 035011.
\bibitem{sata2015}Yohta Sata, Rai Moriya, Takehiro Yamaguchi, Yoshihisa Inoue, Sei Morikawa, Naoto Yabuki, Satoru Masubuchi, and Tomoki Machida, "Modulation of Schottky barrier height in graphene/MoS$_{2}$/metal vertical heterostructure with large current ON–OFF ratio", Japanese Journal of Applied Physics, 2015, 54, 04.
\bibitem{jahangir2017}Ifat Jahangir, M. Ahsan Uddin, Amol K. Singh, Goutam Koley, and M. V. S. Chandrashekhar "Richardson constant and electrostatics in transfer-free CVD grown few-layer MoS$_{2}$/graphene barristor with Schottky barrier modulation >0.6eV",  APPLIED PHYSICS LETTERS, 2017, 111, 142101.
\bibitem{wang2014} Weiyi Wang, Yanwen Liu, Lei Tang, Yibo Jin, Tongtong Zhao and Faxian Xiu,"Controllable Schottky Barriers between MoS$_{2}$ and Permalloy", 2014, 4, 6928.
\bibitem{lili2014} Lili Yu,Yi-Hsien Lee, Xi Ling, Elton J. G. Santos, Yong Cheol Shin, Yuxuan Lin, Madan Dubey, Efthimios Kaxiras, Jing Kong, Han Wang, and Tomas Palacios, "Graphene/MoS$_{2}$ Hybrid Technology for Large-Scale Two-Dimensional Electronics", Nano Lett. 2014, 14, 3055−3063.
\bibitem{zhang2017}Xiuyun Zhang, Zujian Bao, Xiaoshan Ye, Wenxian Xu, Qiang Wang and Yongjun Liu, "Half-metallic properties of 3d transition metal atom-intercalated graphene@MS$_{2}$ (M = W, Mo) hybrid structures", Nanoscale, 2017, 9, 11231–11238.
\bibitem{hai2017}Hai Li, Jiang-Bin Wu, Feirong Ran, Miao-Ling Lin, Xue-Lu Liu, Yanyuan Zhao, Xin Lu, Qihua Xiong, Jun Zhang, Wei Huang,
Hua Zhang, and Ping-Heng Tan, "`Interfacial Interactions in van der Waals Heterostructures of MoS$_{2}$ and Graphene"', ACS Nano 2017, 11, 11714−11723.
\bibitem{erfanifam2017} S. Erfanifam, S.M. Mohseni, L. Jamilpanah, M. Mohammadbeigi, P. Sangpour, S.A. Hosseini, A. Iraji Zad, "Tunable bandgap and spin-orbit coupling by composition control of MoS$_{2}$ and MoO$_{x}$ (x=2 and 3) thin ﬁlm compounds", Materials and Design, 2017, 122, 220–225.
\bibitem{cui2015} Xu Cui, Gwan-Hyoung Lee, Young Duck Kim, Ghidewon Arefe, Pinshane Y. Huang, Chul-Ho Lee, Daniel A. Chenet, Xian Zhang, Lei Wang, Fan Ye, Filippo Pizzocchero, Bjarke S. Jessen, Kenji Watanabe, Takashi Taniguchi, David A. Muller, Tony Low, Philip Kim and James Hone,"'Multi-terminal transport measurements of MoS$_{2}$ using a van der Waals heterostructure device platform"',Nature Nanotechnology, 2015,10.
\bibitem{hua2013} Xiaotian Hua, Xinguo Ma, Jisong Hu, Hua He, Guowang Xu, Chuyun Huang, Xiaobo Chen "Controlling Electronic Properties of MoS$_{2}$/Graphene Oxide Heterojunctions for Enhancing Photocatalytic Performance: the Role of Oxygen", Phys. Chem. Chem. Phys., 2013.
\bibitem{liu2016}Biao Liu, Li-Juan Wu, Yu-Qing Zhao, Ling-Zhi Wang and Meng-Qiu Cai, "First-principles investigation of the Schottky contact for the two-dimensional MoS$_{2}$ and graphene heterostructure", RSC Adv., 2016, 6, 60271.
\bibitem {hieu2017} Nguyen Ngoc Hieu, Huynh Vinh Phuc, Victor V. Ilyasov, Nguyen D. Chien, Nikolai A. Poklonski, Nguyen Van Hieu, and Chuong V. Nguyen,"First-principles study of the structural and electronic properties of graphene/MoS$_{2}$ interfaces",  JOURNAL OF APPLIED PHYSICS, 2017, 122, 104301.
\bibitem{tomer2015}D. Tomer, S. Rajput, L. J. Hudy, C. H. Li,and L. Li, "Carrier transport in reverse-biased graphene/semiconductor Schottky junctions", Appl. Phys. Lett. 106, 173510 (2015).
\bibitem{joon2014} Joon Young Kwak, Jeonghyun Hwang, Brian Calderon, Hussain Alsalman, Nini Munoz, Brian Schutter, and Michael G. Spencer,"Electrical Characteristics of Multilayer MoS$_{2}$ FET’s with MoS$_{2}$ /Graphene Heterojunction Contacts",Nano Lett. 2014, 14, 4511 − 4516.
\bibitem{wan2015}Wen Wan, Xiaodan Li, Xiuting Li, Binbin Xu, Linjie Zhan, Zhijuan Zhao, Peichao Zhang, S. Q. Wu, Zi-zhong Zhu, Han Huang, Yinghui Zhou, and Weiwei Cai, "Interlayer coupling of a direct van der Waals epitaxial MoS$_{2}$/graphene heterostructure", RSC Adv., 2016, 6, 323–330.
\bibitem{rathi2015}Servin Rathi, Inyeal Lee, Dongsuk Lim, Jianwei Wang, Yuichi Ochiai, Nobuyuki Aoki, Kenji Watanabe, Takashi Taniguchi, Gwan-Hyoung Lee, Young-Jun Yu, Philip Kim, and Gil-Ho Kim, "Tunable Electrical and Optical Characteristics in MonolayerGraphene and Few-Layer MoS$_{2}$ Heterostructure Devices", Nano Lett. 2015, 15, 5017−5024.
\bibitem{tomer2017} D Tomer, S Rajput  and L Li, "Spatial inhomogeneity in Schottky barrier height at graphene/MoS$_{2}$ Schottky junctions", J. Phys. D: Appl. Phys., 2017, 50, 165301.
\bibitem{liu2015}Xingen Liu and Zhongyao Li, "Electric field and strain effect on Graphene-MoS$_{2}$ Hybrid Structure: Ab Initio Calculations", J. Phys. Chem. Lett. 2015, 6, 3269−3275.
\bibitem{abbas2014}Abbas Ebnonnasir, Badri Narayanan, Suneel Kodambaka, and Cristian V. Ciobanu, "Tunable MoS$_{2}$ bandgap in MoS$_{2}$-graphene heterostructures", Appl. Phys. Lett., 2014, 105, 031603.
\bibitem{nazir2018}Ghazanfar Nazir, Muhammad Farooq Khan, Sikandar Aftab, Amir Muhammad Afzal,Ghulam Dastgeer, Malik Abdul Rehman, Yongho Seo and Jonghwa Eom, "Gate Tunable Transport in Graphene/MoS2/(Cr/Au) Vertical Field-Effect Transistors",Nanomaterials 2018, 8, 14.
\bibitem{son2017}Da Ye Song, Dongil Chu, Seung Kyo Lee, Sang Woo Pak, and Eun Kyu Kim, "`High photoresponsivity from multilayer MoS$_{2}$/Si heterojunction diodes formed by vertically stacking, J. Appl. Phys., 2017, 122, 124505.
\bibitem{zhao2017}Yudan Zhao, Xiaoyang Xiao, Yujia Huo, Yingcheng Wang, Tianfu Zhang, Kaili Jiang, Jiaping Wang, Shoushan Fan, and Qunqing Li"'Inﬂuence of Asymmetric Contact Form on Contact Resistance and Schottky Barrier, and Corresponding Applications of Diode"', ACS Appl. Mater. Interfaces 2017, 9, 18945−18955.
\bibitem{muatez2012} Muatez Mohammed, Zhongrui Li, Jingbiao Cui and Tar-pin Chen,:"Junction investigation of graphene/silicon
Schottky diodes", Nanoscale Research Letters 2012, 7, 302.
\bibitem{meng2015} Jie Meng, Hua-Ding Song, Cai-Zhen Li, Yibo Jin, Lei Tang, Dameng Liu, Zhi-Min Liao, Faxian Xiu and Da-Peng Yu,"Lateral graphene p–n junctions formed by the graphene/MoS2 hybrid interface"Nanoscale , 2015, 7, 11611.
\bibitem{marco2014} Marco M. Furchi, Dmitry K. Polyushkin, Andreas Pospischil, and Thomas Mueller,"'Mechanisms of Photoconductivity in Atomically Thin MoS$_{2}$", Nano Lett. 2014, 14, 6165−6170.
\bibitem{hyewon2015} Hyewon Du, Taekwang Kim, Somyeong Shin, Dahye Kim, Hakseong Kim, Ji Ho Sung, Myoung Jae Lee, David H. Seo, Sang Wook Lee, Moon-Ho Jo, and Sunae Seo, "'Schottky barrier contrasts in single and bi-layer graphene contacts for MoS$_{2}$ field-effect transistors", Appl. Phys. Lett., 2015, 107, 233106.
\bibitem{huang2017} Yuanyuan Huang, Lipeng Zhu, Zehan Yao, Longhui Zhang, Chuan He, Qiyi Zhao, Jintao Bai, and Xinlong Xu, "Terahertz Surface Emission from Layered MoS$_{2}$ Crystal: Competition between Surface Optical Rectiﬁcation and Surface Photocurrent Surge", J. Phys. Chem. C, 2017.
\bibitem{shih2014} Chih-Jen Shih, Qing Hua Wang, Youngwoo Son, Zhong Jin, Daniel Blankschtein, and Michael S. Strano, "Tuning On-Off Current Ratio and Field-Effect Mobility in a MoS$_{2}$-Graphene Heterostructure via Schottky Barrier Modulation", ACS nano, 2014, 8(6), 5790–5798.
\bibitem{li2016} Yang Li, Cheng-Yan Xu, Jing-Kai Qin, Wei Feng , Jia-Ying Wang, Siqi Zhang, Lai-Peng Ma, Jian Cao, Ping An Hu, Wencai Ren, and  Liang Zhen, "Tuning the Excitonic States in MoS$_{2}$/Graphene van der Waals Heterostructures via Electrochemical Gating", Adv. Funct. Mater. 2016, 26, 293–302.
\bibitem{yue2017}Yuchen Yue, Yiyu Feng, Jiancui Chen, Daihua Zhang and Wei Feng."Two-dimensional large-scale bandgap-tunable monolayer MoS $_{2(1-x)}$ Se$_{2x}$ /graphene heterostructures for phototransistors", J. Mater. Chem. C, 2017, 5, 5887.
\end{thebibliography}
\end{document}